\documentstyle[prl,aps,epsf,prbbib]{revtex}
\newcommand{\intdtw}[1]{\int\!\frac{d^2{\bf #1}}{(2\pi)^2}}

\newcommand{\sumon}{\sum_{\omega_{\nu}}}
\newcommand{\be}{\begin{equation}}
\newcommand{\ee}{\end{equation}}
\newcommand{\rf}[1]{~(\ref{#1})}
\newcommand{\bea}{\begin{eqnarray}}
\newcommand{\eea}{\end{eqnarray}}
\newcommand{\f}{\frac}
\newcommand{\xiq}{\xi({\bf q})}

\newcommand{\Piph}{\Pi_{{\rm ph}}(0)} 
\newcommand{\Pipp}{\Pi_{{\rm pp}}(0)} 
\newcommand{\pa}{\partial}

\def\rv{{\bf r}}
\def\bk{{\bf k}} 
\def\bq{{\bf q}} 
\def\bp{{\bf p}}

\def\bP{{\bf P}}

\def\6{\partial}  \def\b{\beta}

\def\o{\omega}  
  \def\S{\Sigma}
\def\O{\Omega}
 
\def\non{\nonumber\\}

\begin{document}
\draft
\widetext
\twocolumn[\hsize\textwidth\columnwidth\hsize\csname  
@twocolumnfalse\endcsname
\author{P. Pieri, G.C. Strinati, and I. Tifrea\cite{adresa}}
\address{Dipartimento di Matematica e Fisica, Sezione INFM\\ 
Universit\`{a} di Camerino, I-62032 Camerino, Italy\\
}
\date{\today}

\title{Two-dimensional dilute Bose gas in the normal phase}

\maketitle
\hspace*{-0.25ex}

\begin{abstract}
We consider a two-dimensional 
dilute Bose gas above its superfluid transition temperature.
We show that the t-matrix approximation corresponds to 
the leading set of diagrams in
the dilute limit, provided the temperature is sufficiently larger than the 
superfluid transition temperature. Within this approximation, we give an 
explicit expression for the wave vector and frequency dependence of the 
self-energy, and calculate 
the corrections to the chemical potential and the effective mass arising 
from the interaction. We also argue that the breakdown of the t-matrix 
approximation, 
which occurs upon lowering the temperature, provides a simple criterion to 
estimate the superfluid critical temperature for the 
two-dimensional dilute Bose gas.
The critical temperature identified by this criterion coincides with earlier 
results obtained by Popov and by Fisher and Hohenberg using different methods.
Extension of this procedure to the three-dimensional case gives good agreement
with recent Monte Carlo data.
\end{abstract}
\hspace*{-0.25ex}
]

\narrowtext

\section{Introduction}

Renewed interest in two-dimensional (2D) superfluid systems has 
been recently prompted by the discovery of high-temperature 
superconductors. 
Before that, studying 2D bosonic systems was mainly 
motivated by experiments on 
adsorbed helium monolayers\cite{12} and spin-polarized hydrogen recombining 
on a helium film.\cite{21}
More recently, Bose condensation has been achieved experimentally in dilute 
gases of alkali atoms\cite{alk} and in atomic hydrogen,\cite{hyd} 
stimulating active research in this field both experimentally and 
theoretically.\cite{dal}

Even though Bose-Einstein condensation is known not to occur at finite 
temperature for the ideal and interacting boson systems both in one 
and two dimensions, the absence of a condensate 
does not necessarily imply the lack of a phase transition to a 
superfluid state for an interacting 2D Bose system.\cite{14} In this system, 
particles with small momenta behave like a condensate and are responsible 
for the presence of a nonvanishing superfluid density.\cite{pop} 
The critical temperature for the superfluid phase transition was estimated by
Popov\cite{pop} using a functional integral formalism.
The same estimate for the critical temperature was later obtained by Fisher 
and Hohenberg\cite{fis} using a renormalization-group approach, with the result
\be
T_c\approx \f{2\pi n}{m \ln{\ln{\left[1/(nr_0^2)\right]}}} \;\;,
\label{e1}
\ee
where $n$ is the particle density, $m$ the boson mass, and $r_0$ the range
of the interaction potential between bosons (we set $\hbar=k_B=1$ throughout).
In these studies, Popov\cite{pop} and Fisher and Hohenberg\cite{fis} 
approached the phase transition from the superfluid phase and the 
normal phase, respectively. 

The present paper studies the 2D {\em dilute} Bose gas {\em above its 
critical temperature} $T_c$ by relying on conventional diagrammatic methods.
[The criterion for a 2D Bose system to be ``dilute'' will be specified below.]
In this way, results for thermodynamic quantities (like the critical 
temperature and the chemical potential) are most readily obtained. In addition,
these results are amenable to extension to more complex systems, such as the 
composite bosons occurring in the BCS to Bose-Einstein crossover 
problem.\cite{tps2}
Finally, our method enables us to treat on equal footing the dilute Bose gas 
both in two {\em and} three dimensions.

Previous studies of 
the 2D dilute Bose gas have considered the superfluid phase either at zero 
temperature\cite{11,hin,kol,lie} or at finite temperature.\cite{pop,cha} 
At finite temperature, the absence of a condensate in 2D required
either the separation of wave-vector integration into rapid and slow 
parts,\cite{pop} 
or the use of appropriate renormalization group methods.\cite{cha} 
Previous approaches to the 2D dilute Bose gas, however, did not address
the following issues:  
(i) The description of the normal phase (above the critical temperature)
 by standard diagrammatic methods.  
(ii) Upon lowering the temperature, the detection by these methods of a 
superfluid phase transition {\em not} associated with the establishing of 
long-range order.

In this respect, standard many-body diagrammatic methods prove  
sufficient for a complete description of the normal phase of the 2D dilute 
Bose gas. 
In particular, the t-matrix approximation for the self-energy will be shown to
provide the correct description for a dilute Bose gas above $T_c$, akin to the
three-dimensional (3D) case. Both in 2D and in 3D, an explicit analytic 
expression for the self-energy as a function of wave vector and frequency will
be presented.

In our approach, the expression\rf{e1} for the critical temperature obtained 
by Popov\cite{pop} and by Fisher and Hohenberg\cite{fis} appears  
as a {\em lower bound}\/ for the validity of the t-matrix 
as a {\em controlled}\/ approximation for the dilute Bose gas. The 
ensuing diagrammatic classification 
scheme for the dilute Bose gas will, in fact, be shown to break down when the 
temperature $T$ approaches a lower temperature $T_L$, 
which coincides with the estimate\rf{e1} for the 
critical temperature given in Refs.\onlinecite{pop} and\onlinecite{fis}. 
In this way, the occurrence of a superfluid phase transition enters
the diagrammatic theory for the dilute Bose gas, since the physical
mechanism leading to a breakdown of diagrammatic perturbation theory
can only be the presence of a phase transition. 
In addition, by 
applying our method to the 3D case we obtain 
$(T_L-T_{{\rm BE}})/T_{{\rm BE}}\sim n^{1/3} a$, where $T_{{\rm BE}}$ is the
3D Bose-Einstein temperature and $a$ is the scattering length, in 
agreement with recent Monte Carlo simulations\cite{gru,hol} on the 3D dilute 
Bose gas which yield for the critical temperature $T_c$ the same result we 
obtain for $T_L$.

The paper is organised as follows. Section 2 sets up the diagrammatic 
theory for the 2D dilute Bose gas in the normal phase and calculates the 
corrections to the chemical potential and the effective mass due to the 
interaction.   
Section 3 discusses the occurrence of the superfluid phase transition
for the dilute Bose gas through a breakdown of the approximations introduced 
in Section 2. The value of the breakdown-temperature $T_L$ is then given
 both in two and three dimensions.
Section 4 gives our conclusions. 
In the Appendix, the full dependence of the t-matrix self-energy on wave 
vector and frequency is obtained both in 2D and in 3D by exploiting the 
diluteness condition. In addition, the validity of some
approximations on which the theoretical arguments of the text rely is 
explicitly tested numerically.

\section{t-matrix approximation for a dilute system of interacting bosons}
 
In this Section, we analyze the diagrammatic theory for the 2D Bose gas
in the normal phase and determine the leading contributions to the 
self-energy in the dilute limit.
We give an analytic expression for the wave-vector and frequency dependence
of the self-energy, which becomes asymptotically exact in the dilute limit 
(as verified in the Appendix). We further use the zero wave-vector and 
frequency value of this self-energy to dress the single-particle Green's 
functions of the theory in a self-consistent way. This step will enable us
to identify a {\em lower} temperature $T_L$ below which the classification
of diagrams for the dilute Bose system breaks down, as discussed in the next
Section. In addition, we obtain the explicit leading corrections to the 
chemical
potential $\mu$ due to interaction, thus recovering an earlier result by 
Popov\cite{pop}, and to the effective mass. 

We begin by considering a 2D bosonic system interacting via a 
{\em short-range}\/ 
two-body potential $v(\rv)$ with a finite range $r_0$, 
which becomes a $\delta$-function when $r_0\to0$ (cf. Ref.~\onlinecite{kol}). 
We examine the {\em dilute} limit of this system, which is initially 
identified by the condition $n r_0^2\ll 1$, where $n$ is the bosonic density. 
[A stronger condition on the parameter $n r_0^2$ will be required 
below.]
We further 
consider temperatures {\em above} a nominal critical temperature $T_c$ 
but lower than an upper temperature of the order of $T_n\equiv 2\pi n/m$,
at which quantum effects become important.\cite{may}
 
Under these assumptions, the selection of the diagrams yielding the leading 
contributions to the self-energy for the 2D dilute Bose gas in the normal state
proceeds along similar lines as for the 3D dilute Bose gas. 
 
Akin to the 3D case, also in 2D every {\em cycle} (defined as a closed path 
constructed by a sequence of ``bare'' bosonic propagators, with a common 
wave vector and Matsubara frequency flow) introduces a Bose function 
$(e^{(\bq^2/(2m)-\mu)/T}-1)^{-1}$, which appears
after the summation over the common frequency running along the cycle is 
performed.
This function, in turn, cuts off the integral over the remaining wave-vector 
variable approximately at $|\bq|\simeq (m T)^{1/2}$, which is much smaller 
than the cutoff 
$1/r_0$ introduced by the potential (owing to the the diluteness 
condition and the assumption $T\sim T_{n}$). 

The description of the dilute Bose gas accordingly retains only those diagrams
with {\em a minimal number of cycles}, that is, just one cycle. 
These diagrams are shown in Fig.~1(a) and constitute the so-called 
{\em t-matrix approximation} for the self-energy.
\cite{pop,bel}

\begin{figure}
\centering
\epsfxsize=3.1in 
\epsfbox{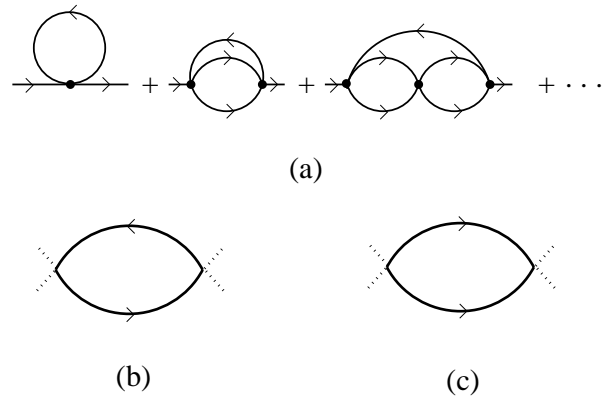}
\vspace{0.4cm}
\caption{(a) t-matrix approximation for the self-energy of an 
interacting Bose gas; (b) Particle-hole bubble; (c) Particle-particle bubble.
}
\label{fig1}
\end{figure}

At $T\sim T_n$ higher-order diagrams can be estimated by replacing the bare 
potential by the t-matrix itself and assigning to each additional cycle a
factor $ t_0\Piph$. Here, $t_0$ and $\Piph$ are the zero frequency and
wave-vector values of the t-matrix and of the particle-hole bubble (Fig.~1(b)),
respectively. It will further shown below that the product 
$t_0\Piph\propto 1/\ln[1/(n r_0^2)]\ll 1$ in the dilute limit.
   
We will verify, however, that the classification of diagrams
based on the cycle argument breaks down upon lowering the temperature
down to a value $T_L$. More precisely, we will find that the t-matrix 
correctly describes the dilute Bose gas in the temperature range
$T_L\lesssim T\lesssim T_{n}$, in the sense that no other diagrams 
besides the t-matrix itself need to be included. 
 
The self-energy corresponding to the set of diagrams depicted in Fig.~1(a)
reads: 
\bea
& &\Sigma_t(q)=- T \sum_{\omega_{\nu'}}\int\!\f{d^2\bq'}{(2\pi)^2} 
 G(q')\left[ t\left(\f{\bq'-\bq}{2},\f{\bq'-\bq}{2},q+q'\right)\right.\non
& &+ t\left.\left(\f{\bq-\bq'}{2},
\f{\bq'-\bq}{2},q+q'\right)\right]
\label{sel}
\eea
where the t-matrix $t(\bp,\bp',P)$ is defined by the integral 
equation
\bea
t(\bp,\bp',P)&=&v(\bp-\bp') -
T \sum_{\omega_{\nu}}\int \f{d^2\bq}{(2\pi)^2}\; 
v(\bp-\bq)\non
&\times&G\left(\frac{P}{2}-q\right)G\left(\frac{P}{2}+q\right)t(\bq,\bp',P)
\label{tma}
\eea
with the notation $q\equiv(\bq,i\omega_{\nu})$ 
($\omega_{\nu}=2\pi\nu T$ - $\nu$ integer - being a bosonic Matsubara 
frequency).~\cite{sym}

To lowest order in the density, all single-particle Green's functions $G$ 
in Eqs.\rf{sel} and\rf{tma} are considered to be bare ones. Quite generally,
self-energy insertions in the Green's functions become relevant 
when the phase transition is approached.
It is shown in the Appendix that, in the dilute limit, the dependence of
the self-energy $\S_t(q)$ on $q$ can be disregarded when calculating physical 
quantities, so that one may set $\S_t(q)\simeq\S_t(0)$ in all single-particle
Green's functions entering Eqs.\rf{sel} and \rf{tma} and re-absorb the constant
$\S_t(0)$ by a shift of the chemical potential. In this way, when the chemical
potential is expressed in terms of $n$ and $T$, one obtains
\be
\mu-\S_t(0)=\mu_0 
\label{mu0}
\ee
where $\mu_0=\mu_0(n,T)$ is the chemical potential of the 2D {\em ideal} Bose
gas.\cite{mub} 

In the following, we shall consider all Green's functions in Eqs.\rf{sel} 
and\rf{tma} to be self-consistently dressed by the 
self-energy\rf{sel} with $q=0$, resulting in an ``improved'' t-matrix 
approximation which can be adopted to approach the critical 
temperature more closely. 

The explicit expression of $\S_t(q)$ for arbitrary values of $q$ is given by 
Eq.\rf{selan2} of the Appendix, which is asymptotically valid in the dilute 
limit. 
From that 
expression, the shift of the chemical potential due to interaction as well as
the relevant effective mass can be obtained. For the chemical potential, it is
sufficient to know the value of $\S_t(0)$, which from Eq.\rf{selan2} 
becomes:\cite{asy}
\be
\Sigma_t(0)\approx\frac{8 \pi n}{m \ln \left[1/(m |\mu_0| 
r_0^2)\right]}
\label{sigma0}
\ee
By entering in this expression the analytic form of the chemical 
potential $\mu_0$,\cite{mub} we obtain eventually
\bea
\Sigma_t(0) &\approx &\frac{8 \pi n/m}
{\ln (1/[m T r_0^2 |\ln (1-e^{-T_n/T})|])}\non
&\approx&\frac{8\pi n/m}{\ln[1/(n r_0^2)]+\ln[T_n/(2\pi T)]
-\ln|\ln(1-e^{-T_n/T})|}\non
&\approx&\frac{8 \pi n}{m \ln \left[1/(n r_0^2)\right]}
\label{si}
\eea
the last result holding for $T_n/\ln[1/(n r_0^2)]\ll T\lesssim 
T_{n}$, which includes the temperature range ($T_L\lesssim T \lesssim T_{n}$)
of physical interest. 
Equation\rf{si} provides the leading self-energy term for the 2D dilute Bose 
gas in the normal state, from which the shift of the chemical potential is
obtained as $\mu=\mu_0+\Sigma_t(0)$. 

Note that the expression\rf{si} is temperature independent in the temperature
range $T_L\lesssim T \lesssim T_n$ we are considering. The only 
temperature dependence of $\mu$ thus originates from $\mu_0$. In 
particular, at $T=T_c$ we substitute the
value of $T_c$ from Eq.\rf{e1} onto the expression\rf{si}, yielding:
\bea
\mu(T_c)&=&\mu_0(T_c)+\Sigma_t(0)\non
&\approx&-\frac{2 \pi n}{m \ln \left[1/(n r_0^2)\right]}
\frac{1}{\ln\ln[1/(n r_0^2)]}+\frac{8 \pi n}{m \ln \left[1/(n r_0^2)\right]}
\non
&\approx&\frac{8 \pi n}{m \ln \left[1/(n r_0^2)\right]} \;\; ,
\label{mutc}
\eea 
provided that $\ln\ln(1/(n r_0^2)$ is sufficiently larger than unity (dilute 
limit).
This result coincides with the value of the chemical potential
obtained in Refs.~\onlinecite{pop,fis} where 
the critical temperature was approached from below. 

The effective mass can eventually be calculated after analytic continuation of
$\Sigma_t(q)$. 
The result is
\be
\f{m^*}{m}=1+\frac{1}{4\ln\ln[1/(n r_0^2)]}\;.
\label{mass}
\ee 
Details of the derivation of Eq.\rf{mass} are reported in the Appendix.
\section{Breakdown of the bosonic t-matrix approximation and the superfluid
phase transition}

We pass now to show that the selection of diagrams made in the previous 
Section, which was based on the cycle argument, breaks down upon lowering the
temperature when the superfluid phase transition is approached.
Specifically, consideration of the temperature at which 
the particle-hole diagrams
(which where discarded by the cycle argument) are no longer 
negligible in comparison with the particle-particle diagrams, will lead us 
to identify a {\em lower}\/ temperature $T_L$ that turns out to coincide 
with the critical temperature\rf{e1} determined in Refs.\onlinecite{pop}
and\onlinecite{fis}.

To determine the range of validity of the t-matrix approximation, it is 
enough to compare the particle-particle bubble $\Pi_{\rm{pp}}$ of Fig.~1(c)
(which constitutes the building block of the t-matrix of Fig.~1(a)) with the 
{\em particle-hole bubble} $\Pi_{\rm{ph}}$ of Fig.~1(b).
This statement follows from the classification of diagrams we have made 
because $t_0\simeq \Pipp^{-1}$ in the dilute limit [cf.~Eqs.\rf{tap} 
and\rf{tmapp}], yielding $\Piph/\Pipp$ as the small parameter of the theory at 
$T\sim T_n$. The ratio $\Piph/\Pipp$ grows, however, upon lowering the 
temperature below $T_n$ and reaches the value of unity at the temperature 
$T_L$ introduced above.

The {\em particle-particle bubble} is given by
\be
\Pi_{{\rm pp}}(0)=T\sum_{\omega_{\nu}}\int^{r_0^{-1}} \!\f{d^2\bq}{(2\pi)^2} \;
 G(q) \; G(-q)
\label{e6}
\ee
where 
$
G(q)=(i\omega_{\nu} -\bq^2/(2m) +\mu_0)^{-1} 
$  
according to the arguments above.
With the notation $\xi(\bq)=|\bq|^2/(2m)-\mu_0$, we obtain:
\bea
\Pi_{{\rm pp}}(0)&=&\int^{r_0^{-1}} \!\f{d^2\bq}{(2\pi)^2}\; 
\frac{1+2 n_{B}(\xiq)}{2\xiq}
 \non
&\approx&\frac{m}{4\pi}\ln \left(\f{1}{2 m |\mu_0| r_0^2}\right)
+ \frac{1}{2\pi}\int_0^{r_0^{-1}} d |\bq|\; \frac{|\bq| \, n_B(\xiq)}{\xiq} 
\non
&\approx&\frac{m}{4\pi}\ln \left(\f{1}{2 m |\mu_0| r_0^2}\right)  
\label{pp1}
\eea
where $n_B(x)=(e^{x/T}-1)^{-1}$ is the Bose function and the last 
asymptotic 
equality holds in the dilute limit $n r_0^2\ll 1$ and for 
temperatures of the order of $T_n$ (such that $|\mu_0|\sim T_n$).

The {\em particle-hole bubble} for $q=0$ is, as usual, given by 
$\Piph=\6 n/\6\mu$. 
For the two dimensional Bose gas the chemical potential is known analytically
for all temperatures and densities\cite{mub}, such that
\be
\f{\6 n}{\6 \mu}=\f{m}{2\pi}\f{e^{\mu_0/T}}{1-e^{\mu_0/T}} \; .
\label{com}
\ee
At $T\sim T_n$, $\mu_0(n,T_n)\sim T_n$ and $\6 n/\6\mu\sim m$, such that the 
ratio $\Piph/\Pipp\approx 1/\ln[1/(n r_0^2)]\ll 1$ in the dilute limit.
When $T\ll T_n$, on the other hand, $|\mu_0|\ll T$ and 
$\6 n/\6 \mu\approx m T/(2\pi|\mu_0|)$, which is much larger than the 
corresponding value at $T\sim T_n$. A lower temperature $T_L$ can thus be 
reached, such that the ratio $\Piph/\Pipp$ equals unity when
\be
\f{1}{2}\ln\f{1}{2 m |\mu_0| r_0^2}=\f{T_L}{|\mu_0|}\; .
\label{conf}
\ee
Entering now the asymptotic expression $|\mu_0|\approx T e^{-T_n/T}$, which holds 
for  $T\ll T_n$, yields eventually
\be
T_L\approx\f{T_n}{\ln\ln[1/(n r_0^2)]}
\ee  
valid under the assumption $\ln\ln[1/(n r_0^2)]\gg 1$ (which 
{\em defines} the diluteness condition in 2D). This expression for 
$T_L$ coincides with the estimate\rf{e1} for the critical temperature given 
by Popov\cite{pop} and by Fisher and Hohenberg\cite{fis}. 
 Note that the {\em double-log
 dependence} of
$T_L$ on $n r_0^2$ originates, on the one hand, from the log dependence of
$\Pipp$ on $|\mu_0|$ and, on the other hand, from the exponential dependence
of $\mu_0$ on $T$ at low temperatures. Note also that the more stringent 
diluteness condition $\ln\ln[1/(n r_0^2) \gg 1$ (in the place of the 
original $n r_0^2 \ll 1$) 
is required to get a {\em finite} temperature range 
($T_L\lesssim T \lesssim T_n$), where self-energy diagrams can be selected by 
the diluteness condition. 

On physical grounds, the only mechanism which may lead to 
the breakdown of the diagrammatic classification in terms of a small 
parameter (as explicitly seen above) is {\em the occurrence
of a phase transition}. The breakdown temperature $T_L$ is thus 
expected to provide an estimate for the superfluid critical temperature $T_c$,
also because the two temperatures $T_L$ and $T_c$ are expected to have the same
functional dependence on $n r_0^2$, 
as comparison of our expression for $T_L$ with the value\rf{e1} for the 
critical temperature obtained in Refs.\onlinecite{pop,fis} indeed shows.

Our identification of $T_L$ with $T_c$ is further confirmed by applying    
the same sort of arguments to the 3D dilute Bose gas, for which Monte Carlo
results are available.\cite{gru,hol} In this case, the 
particle-particle and particle-hole bubbles are given, respectively, by
\be
\Pi_{{\rm pp}}(0)\approx\frac{m}{2 \pi^2 r_0}=\f{m}{4\pi a}
\ee
where $a$ is the scattering length, and
\be
\Pi_{{\rm ph}}(0)=
\f{\6 n}{\6 \mu}\approx\f{2 m n^{1/3}}{3 \zeta(3/2)^{4/3}} 
\f{T_{\rm BE}}{T-T_{\rm BE}}
\ee
since $\mu_0\simeq -9\zeta(3/2)^2 (T-T_{{\rm BE}})^2/(16 \pi T_{\rm{BE}})$ 
for the chemical potential of the ideal Bose gas (valid near the Bose-Einstein
temperature $T_{{\rm BE}}=2\pi/\zeta(3/2)^{2/3} n^{2/3}/m$).
The temperature at which these contributions coincide then 
defines the 3D breakdown temperature $T_L^{3D}$. One obtains:
\be 
\f{T_L^{3D} -T_{{\rm BE}}}{T_{{\rm BE}}}= \f{8 \pi}{3 \zeta(3/2)^{4/3}} 
n^{1/3} a \simeq 2.33 \,n^{1/3} a\;.
\label{t3d}
\ee
This result agrees with recent Monte Carlo simulations for
the 3D hard-core Bose gas in the dilute limit,\cite{hol} which yielded
precisely $(T_c-T_{{\rm BE}})/T_{{\rm BE}}=(2.3 \pm 0.25) n^{1/3} a$, and 
coincides with the analytic result of Ref.\onlinecite{bay2}, which was 
obtained by a completely different method.
 
The result\rf{t3d} should be regarded as altogether non trivial, since 
previous analytic treatments of the 3D
dilute Bose gas resulted either in different dependences of 
$(T_c- T_{{\rm BE}})/T_{{\rm BE}}$ on the parameter
$n^{1/3} a$, e.g., of the type $(n^{1/3} a)^{1/2}$ [cf.~Ref.~\onlinecite{hua}]
or $(n^{1/3} a)^{2/3}$ [cf.~Ref.~\onlinecite{sto1}], or in the same linear 
dependence on the parameter $n^{1/3} a$, but with a different 
proportionality coefficient.\cite{lee,sto,bay}
In our approach, the linear dependence on the parameter $n^{1/3} a$ of the 
temperature shift has been directly related to the quadratic dependence of 
the free-boson chemical potential on $T-T_{{\rm BE}}$ near $T_{{\rm BE}}$ in 
3D. 

\section{Concluding remarks}

In this paper, we have considered the two-dimensional dilute Bose gas in the
normal phase, in the interesting temperature region ranging from an 
{\em upper} temperature $T_{n}$ (below which quantum effects become important) 
to a {\em lower}
temperature $T_L$ (which we have identified as the superfluid critical 
temperature). In this temperature region we have 
analyzed the ordinary diagrammatic theory
and organized it in powers of the parameter $1/\ln[1/(n r_0^2)]$, which was 
assumed to be small compared to unity.

In this way, the standard t-matrix has been identified as yielding the
dominant set of diagrams for the self-energy when $n r_0^2\ll 1$. 
Further analysis
of the theory to define the temperature range where the t-matrix 
approximation holds has, however, led us to consider the stronger condition 
$\ln \ln [1/(n a^2)]\gg 1$ as characteristic of the ``dilute'' Bose gas in two 
dimensions, thus confirming the criterion introduced by Fisher and 
Hohenberg\cite{fis} via different methods.

Our identification of the lower temperature $T_L$ (and, thus, of the superfluid
critical temperature) rests on the finding that the diagrammatic 
classification scheme for the 2D dilute Bose gas breaks down at this lower 
temperature, in the sense that additional diagrams (besides the t-matrix) 
become also important at $T_L$ and the hierarchy established
for the dilute gas no longer holds. 
In this respect, it may be worth mentioning that our criterion to identify the
critical temperature does not contradict the usual criterion which defines
$T_c(n)$ as the temperature where the equation $\mu(n,T_c)=\Sigma(q=0; n,T_c)$
is satisfied.
Solving, in fact, for this equation requires one to rely on an approximation 
for the self-energy which is valid {\em even at} $T_c$. We have seen, however, 
that in our case the t-matrix approximation for the self-energy breaks down
{\em before} reaching $T_c$, as soon as the critical region above $T_c$ is 
approached. 

The finding that an estimate of the superfluid critical temperature can be
obtained from the ordinary diagrammatic theory in the normal phase, both in
two and three dimensions, constitutes {\em per se}\/ a nontrivial result, 
especially because the nature of the (superfluid) 
transition in two and three dimensions is quite different 
(involving, respectively, quasi-long-range order and 
true long-range order). In addition, 
our approach is rather straightforward and amenable to direct implementation 
to more complex physical situations. 

In this respect, a possible application of our results may be the normal 
state of high-temperature cuprate superconductors, which are 
quasi-two-dimensional systems. Experiments related to the 
normal\cite{2,3} and superconducting\cite{1} state in these systems suggest,
in fact, that a correct description of their properties 
might require an intermediate (crossover) approach between the 
Fermi liquid theory (weak-coupling) and the dilute Bose gas approach 
(strong-coupling). 
For this reason, crossover theories have been considered by several authors, 
both for the normal state\cite{7,8} and the broken-symmetry phase.\cite{5,6} 
Since a reliable study of the crossover problem should require a good 
knowledge of at least the extreme (weak- and strong-coupling) limits,
the approach developed in this paper might shed light on the 
strong-coupling limit of these theories.\cite{tps2}

\acknowledgments

We are indebted to C. Castellani, P. Nozi{\`e}res, and F. Pistolesi for 
helpful discussions.  
I. T. gratefully acknowledges financial support from the Italian INFM under 
contract No. PRA-HTSC/96-99.

\appendix
\section{}
In this Appendix, we obtain an analytic expression for the t-matrix 
self-energy, which is valid above $T_L$ in the dilute limit. From this 
expression, the effective mass (at $\bq=0$) is calculated.
In addition, the approximation
\be 
\Sigma_t(q)\simeq\Sigma_t(0)
\label{cru}
\ee
(which is crucial for the arguments presented in the text) will be justified 
theoretically and checked numerically. Both two-dimensional and 
three-dimensional cases will be considered for completeness.
\subsection{Two dimensions}
It is convenient to parametrize the short-range potential by a separable 
potential in wave-vector space, by setting $v(\bk-\bk')=
v_0 w_{\bk}w_{\bk'}$ with $w_{\bk}=\theta(k_0-|\bk|)$ and $k_0=r_0^{-1}$.
In this way, Eq.\rf{tma} can be readily solved to yield 
$t(\bp,\bp',P)=w_{\bp} w_{\bp'} t(P)$, with
\be
t^{-1}(P)=v_0^{-1}+ T\sumon\intdtw{q}\, w_{\bq}^2\; G\left(\f{P}{2}-q\right)
G\left(\f{P}{2}+q\right)\;, 
\label{tap}
\ee 
where $G(q)=(i\omega_{\nu}-\bq^2/2m +\mu_0)^{-1}$, consistently with the 
assumption\rf{cru} and the ensuing Eq.\rf{mu0}. This choice of $G(q)$ will 
lead us to verify the key assumption\rf{cru} in a self-consistent manner. 
The frequency sum in Eq.\rf{tap} can be performed explicitly, yielding
\bea
& &T\sumon\intdtw{q}\; w_{\bq}^2\; G\left(\f{P}{2}-q\right)G\left(\f{P}{2}
+q\right)\non
& &=\intdtw{q}w_{\bq}^2\;
\frac{1+n_B(\xi_{\bP/2-\bq})+n_B(\xi_{\bP/2+\bq})}{\bP^2/(4m)+\bq^2/m-2\mu_0-i
\Omega_{\nu}}\;.
\eea
In the dilute limit ($n r_0^2\ll 1$) and for $T_L\lesssim T\lesssim T_n$, the 
Bose functions appearing in the above expression can be neglected, since they 
yield contributions smaller by a factor $T r_0^2\ll 1$ with respect to the 
term retained.
The integration over the wave vector $\bq$ then yields
\bea
t^{-1}(P)&=&\f{1}{v_0}\non
&+&\frac{m}{4\pi}\ln\left[\f{k_0^2/m + 2|\mu_0| +\bP^2/(4m) -i
\O_{\nu}}{ 2|\mu_0| +\bP^2/(4m) -i\O_{\nu}}\right]\label{tma2}
\label{tmapp}
\eea
where ln stands for the principal branch of the complex logarithm.
The t-matrix self-energy is obtained by inserting expression\rf{tmapp}
into Eq.\rf{sel} which, for a separable potential, becomes
\bea
\Sigma_t(\bq,\omega_{\nu})&=&-2 \;T\sum_{\O_{\nu'}}\intdtw{q'}\; w^2_{(\bq'-\bq)/2}\non 
&\times& G(\bq',\O_{\nu'}-\o_{\nu})  t(\bq+\bq',\Omega_{\nu'})\;.
\label{selta}
\eea
To perform the frequency sum in Eq.\rf{selta} we exploit the
analytic properties of $t(\bP,z)$. From Eq.\rf{tmapp},
after the replacement $i\Omega_{\nu'}\to z$, it can be readily verified that 
$t(\bP,z)$ has a simple pole for $z=\bP^2/(4 m) + 2 |\mu_0| + k_0^2/[m
(1-e^{-1/{\tilde v_0}})]$ (with ${\tilde v_0}=m v_0/(4\pi)$) and 
a branch cut along the real axis for 
$\bP^2/(4 m) + 2 |\mu_0|< {\rm Re} (z)<\bP^2/(4 m) +2 |\mu_0|+k_0^2/m$.
The frequency sum in Eq.\rf{selta} can be then performed by a contour 
integration, yielding three distinct contributions: one from
the simple pole of the Green's function $G(\bq',z)$, one from the 
simple pole of the t-matrix, and one from the integration 
along the cut of $t(\bq+\bq',z)$. The term originating from the simple pole of
the t-matrix, which occurs for ${\rm Re}(z)> k_0^2/m$, is exponentially 
suppressed by the Bose factor $1/(\exp(\beta z)-1)$ and is thus negligible.  
The term associated with the integration along the cut can be estimated to be
smaller than the term from the simple pole of the Green's 
function by a factor $1/\ln[1/(n r_0^2)]$, and is also negligible in the 
dilute limit.
We are thus left with the expression:
\bea
& &\Sigma_t(\bq,\omega_{\nu})=2 
\intdtw{\bp} \f{w^2_{\bq/2}}
{e^{\b [\f{(\bq -\bp)^2}{2m}+|\mu_0|]}-1}\left[\f{1}{v_0}+\f{m}{4\pi}\phantom{
\f{k_0^2/m}{\bq^2/m}}\right. 
\non
&\times &\left.
\ln\f{k_0^2/m+|\mu_0|+\bq^2/(2m)-\bp^2/(4m)-i\omega_{\nu}}{|\mu_0|+\bq^2/(2m)
-\bp^2/(4m)-i\omega_{\nu}}\right]^{-1}.
\label{temp}
\eea
Note that the Bose factor in Eq.\rf{temp} is peaked about
$\bp=\bq$ with a width of order $T^{1/2}$, which is smaller than 
the range of $\bp$ over which the log-term in the denominator of Eq.\rf{temp}
varies appreciably. When performing the integration over $\bp$ in 
Eq.\rf{temp}, we can then approximate $\bp=\bq$ in the logarithm, which is in 
this way factored out of the integral, yielding the following asymptotic 
expression for the self-energy:
\bea
\f{\S_t(q)}{2 n}&\approx&w^2_{\bq/2}\left\{ 
\f{1}{v_0}\phantom{\f{k_0^2/m}{q^2}}\right.\non
&+&\left.\f{m}{4\pi}\ln\left[\f{k_0^2/m + |\mu_0| +\bq^2/(4m) 
-i\o_{\nu}}{ |\mu_0| +\bq^2/(4m) -i\o_{\nu}}\right]\right\} ^{-1}.
\label{selan2} 
\eea

The analytic expression\rf{selan2} can be compared with the numerical 
calculation for the t-matrix self-energy, obtained by retaining all 
contributions to Eq.\rf{selta}.

\begin{figure}
\centering
\epsfxsize=3.1in 
\epsfbox{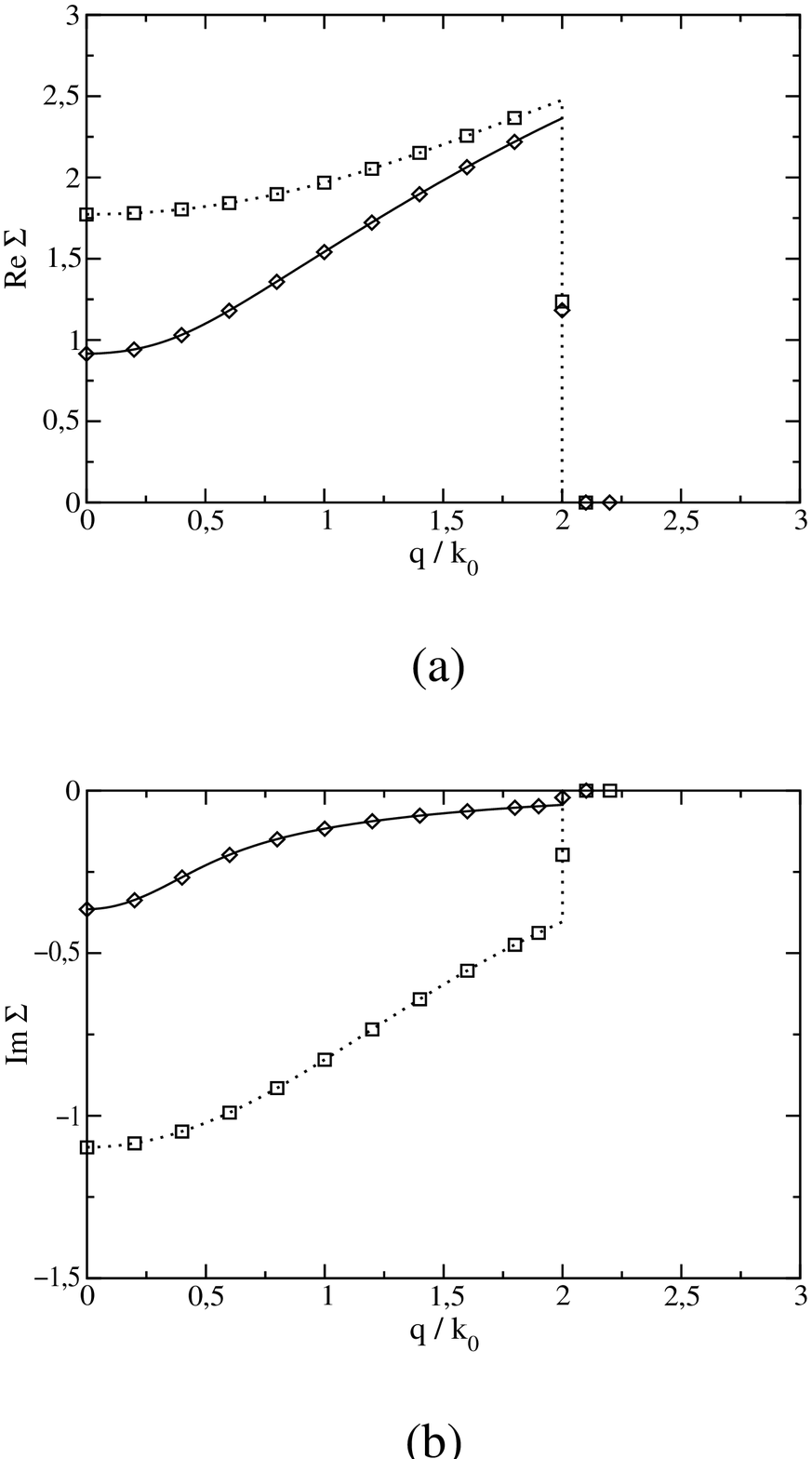}
\vspace{0.4cm}
\caption{ $\S(q,\omega_{\nu})$ (units of $T_n$) for a 2D dilute Bose gas
at $T=T_n$ and $n r_0^2=1.6\times 10^{-3}$. 
Numerical results for $\nu=10$ (diamonds) and $\nu=100$ (squares) are 
compared with the analytic expression\rf{selan2} (full and dotted lines,
respectively).  
}
\label{fig2}
\end{figure}

We have found that the asymptotic expression\rf{selan2} 
reproduces extremely well the numerical results for\rf{selta} when the 
diluteness parameter is sufficiently small. As an example, in Fig.~2 the 
analytic expression for $\Sigma_t(q)$ is 
compared with the numerical results for the choice of parameters $T=T_n$, 
$m v_0/(4\pi)=1$, and $n r_0^2= 1.6\times 10^{-3}$. One can see that 
in this case the agreement is excellent.

The expression\rf{selan2} for the self-energy can also be used to 
verify that, when evaluating physical quantities (such as the density $n$),
the wave-vector and frequency dependence of $\S_t(q)$ is actually irrelevant,
so that one can approximate $\Sigma_t(q)\simeq\Sigma_t(0)$ as anticipated in
Eq.\rf{cru}.
This is because the presence of the logarithm in Eq.\rf{selan2} makes the 
dependence 
of $\S_t(q)$ on $q$ rather slow. The approximation $\S_t(q)\simeq\S_t(0)$ is
thus
justified over a large portion of $q$ space and can be exploited to evaluate 
physical quantities. In particular, for $T=T_n$ we have obtained numerically 
that the relative error when evaluating $n=- T\sum_{\nu}\int \f{d^2\bq}
{(2\pi)^2}\; G(q)$ alternatively with $G(q)=[i\O_{\nu} -\bq^2/(2 m)+\mu-
\S_t(q)]^{-1}$ or with $G(q) \to G_0(q)=[i\O_{\nu} -\bq^2/(2 m)+\mu-
\S_t(0)]^{-1}$ is less than 1\% for $n r_0^2\lesssim 10^{-2}$. 
  
Finally, the t-matrix self-energy\rf{selan2} can be exploited to calculate the 
effective mass for the dilute Bose gas.
Recall that, once the retarded self-energy is known, 
the effective mass can be calculated as\cite{neg}
\be
\f{m^*}{m}=\left(1-\f{\pa {\rm Re}\S(|\bq|,\o)}{\pa \o}\right)
\left[1+ \f{m}{|\bq|}\f{\pa {\rm Re}\S(|\bq|,\o)}{\pa |\bq|}\right]^{-1},
\label{effm}
\ee
where the derivatives are meant to be calculated at the quasi-particle pole
defined by the equation
\be
\o_{\bq}=\f{\bq^2}{2m}-\mu+ {\rm Re}\S(\bq,\o_{\bq})\;.
\label{pole}
\ee
In general, the effective mass $m^*$ depends on $\bq$. Here we are
interested in its value at $\bq=0$, which is relevant at low 
temperatures. 

The self-energy\rf{selan2} can be analytically continued via the 
replacement $i \omega_n\to \omega + i 0^+$. The quasi-particle-pole 
equation\rf{pole} at $\bq=0$ can then be solved asymptotically, to yield
\be
\o_0=-\mu+\S_t(0)\left(1 -2 \f{\ln\ln[1/(n r_0^2)]}{\ln(1/(n r_0^2)]}\right) 
\label{om0}
\ee
as it can be verified by inserting the value\rf{om0} for $\o_0$ in the 
quasi-particle-pole equation and by discarding subleading terms in the dilute
limit $n r_0^2\ll 1$.
The derivatives of $\Sigma(\bq,\o)$ at $(\bq=0,\o=\o_0)$ can also be readily
calculated. The effective mass at $\bq=0$ is then given by
\be
\f{m^*}{m}\approx\f{1+1/(2\ln\ln[1/(nr_0^2)])}{1-1/(4\ln\ln[1/(nr_0^2)])}
\approx 1+1/(4\ln\ln[1/(nr_0^2)]).
\ee                                                        
Note the occurrence of the same double-log dependence characteristic of the
temperature $T_L$.

\subsection{Three dimensions}
The three-dimensional case can be treated in a parallel fashion to the 
two-dimensional case. By considering the same separable potential adopted
in 2D, and by following the same steps which lead to Eq.\rf{tma2}, we 
now obtain:
\bea
t^{-1}(P)&=&\f{1}{v_0}+\f{m}{2\pi^2}\left[ k_0-m^{1/2}\sqrt{- 2\mu_0 +
\f{\bP^2}{4m} -i\O_{\nu}}\right.
\non
&\times &\left. \arctan\f{k_0/m^{1/2}}
{\sqrt{- 2\mu_0 +\f{\bP^2}{4m} -i\O_{\nu}}}\right]\label{tma3}
\eea
where the complex arctan is defined, as usual, in terms of the principal 
branch of the complex logaritm as follows:
\be
\arctan z =\f{i}{2}\ln\f{1- i z}{1+ i z}\;.
\ee
Like in 2D, the t-matrix $t(\bP,z)$ has a branch cut along the real axis for
$\bP^2/(4 m) + 2 |\mu_0|< {\rm Re} (z)<\bP^2/(4 m) +2 |\mu_0|+k_0^2/m$, 
and a simple pole located along the real axis for 
${\rm Re} z> \bP^2/(4 m) +2 |\mu_0|+k_0^2/m$. Upon trasforming the 
frequency sum in\rf{selta} into a contour integration, the t-matrix simple 
pole contribution will be again strongly suppressed by the Bose factor, while 
the term associated with the integration along the cut can now be proven
to be smaller than the term originating from the simple pole of
the Green's function by a factor $(n r_0^3)^{1/2}$.
In the dilute limit, only the simple pole of the Green's function therefore 
contributes to the contour integration. By the same argument leading to
Eq.\rf{selan2}, we obtain the following asymptotic expression for the
t-matrix self-energy:
\bea
\f{\S_t(q)}{2 n}&\approx&w^2_{\bq/2}\left[
\f{1}{v_0}+\f{m}{2\pi^2}\left(k_0 - \sqrt{|\mu_0| +\f{\bq^2}{4m} 
-i\o_{\nu}}\right.\right. \non
&\times & \left.\left. m^{1/2}\arctan\f{k_0/m^{1/2}}
{\sqrt{ |\mu_0| +\f{\bq^2}{4m} -i\o_{\nu}}}\right)\right]^{-1}\;.
\label{selan3}
\eea

The asymptotic expression\rf{selan3} has also been checked against numerical
calculation of Eq.\rf{selta} in three dimensions. In Fig.~3 the analytic 
expression\rf{selan3} is compared with the
numerical results for the choice of parameters $T=T_n$, 
$(2\pi^2)/(m v_0 k_0)\to 0$, and $n^{1/3} r_0= 1 \times 10^{-2}$. Even in 3D 
the agreement is excellent.

The approximation $\S_t(q)\simeq\S(0)$ has further been checked by evaluating
the particle density. In this case, we have found that the error introduced
by the approximation\rf{cru} in the estimate for the density is less than 1\% 
when $n r_0^3\lesssim 5\times 10^{-3}$. 
 
Finally, the solution of the quasi-particle-pole equation\rf{pole} at $\bq=0$ 
in 3D is given by
\be
\o_0=-\mu+\S_t(0)=|\mu_0|\;,
\ee 
while the retarded self-energy near the quasi-particle pole is given by
\be
\S(\bq,\o)=\f{2 n}{\f{m}{4\pi a} - i \f{m}{2\pi^2}A(\bq,\o)
\left[\f{\pi}{2}+\f{i}{2}\ln\f{\pi/(2 a A(\bq,\o))+1}{\pi/(2 a A(\bq,\o))-1}
\right]}
\ee 
with $A(\bq,\o)=\sqrt{\o-\bq^2/(4 m)-|\mu_0|}$ and where we have introduced 
the scattering length $a$ via the relation $m/(4\pi a)\approx 
(m k_0)/(2\pi^2)$.
By using the definition\rf{effm}, the effective mass at $\bq=0$ can be 
readily calculated, leading to the result:
\bea
\f{m^*}{m}&\approx&\left[1+\f{8\pi n a^3}{m}\left(1+\f{4}{\pi^2}\right)\right]
\left[1+\f{4\pi n a^3}{m}\left(1+\f{4}{\pi^2}\right)\right]^{-1}\non
&\approx& 1+\f{4\pi n a^3}{m}\left(1+\f{4}{\pi^2}\right)\;\;.
\eea  
Note that the interaction increases the (quasi)-particle mass both in 3D and
in 2D with respect to its bare value.

\begin{figure}
\centering
\epsfxsize=3.1in 
\epsfbox{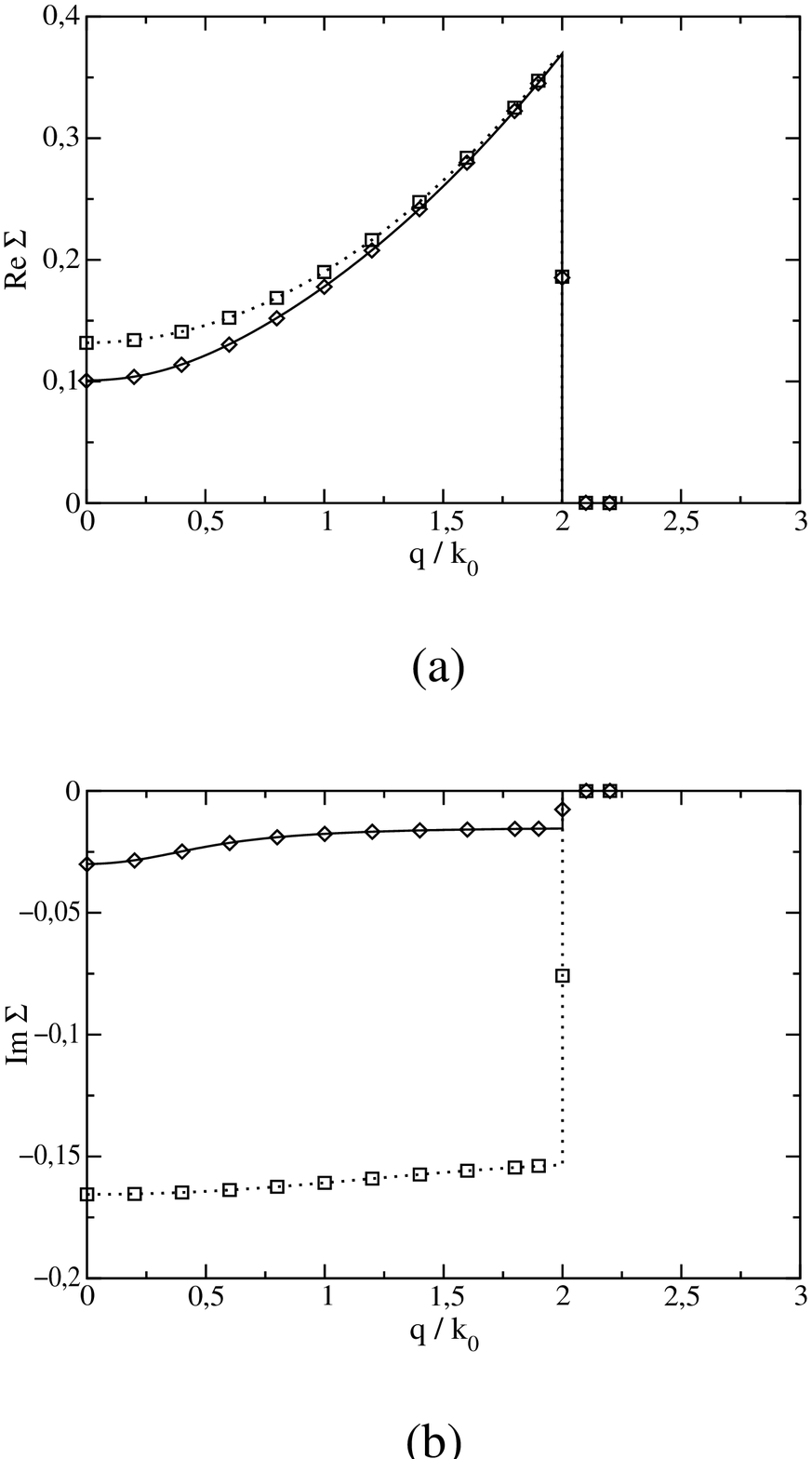}
\vspace{0.4cm}
\caption{ $\S(q,\omega_{\nu})$ (units of $T_n$) for a 3D dilute Bose gas
at $T=T_n$ and $n^{1/3} r_0=1\times 10^{-2}$. 
Numerical results for $\nu=10$ (diamonds) and $\nu=100$ (squares) are compared
with the analytic expression\rf{selan3} (full and dotted lines, respectively).
}
\label{fig3}
\end{figure}



\begin{references}
\bibitem[*]{adresa} Permanent address: Dept. of Theoretical 
Physics, ``Babes-Bolyai'' University, 3400 Cluj, Romania.
\bibitem{12} W.D. McCornick, D.L. Goodstein, and J.G. Dash, Phys. Rev. 
{\bf 168}, 249 (1968); G.A. Stewart and J.G. Dash, Phys. Rev. A {\bf 2}, 918
(1970); J.G. Dash, Phys. Rep. {\bf 38 C}, 177 (1978); D.J. Bishop and J.D. 
Reppy, Phys. Rev. B {\bf 22}, 5171 (1980).
\bibitem{21} L.J. Lantto and R.M. Nieminen, J. Low Temp. Phys. {\bf 37}, 1 
(1979).
\bibitem{alk}
M.H. Anderson {\em et al.}, Science {\bf 269}, 198 (1995);
C.C. Brabley {\em et al.}, Phys. Rev. Lett. {\bf 75}, 1687 (1995);
K.B. Davis {\em et al.}, Phys. Rev. Lett. {\bf 75}, 3969 (1995).
\bibitem{hyd}
D.G. Fried {\em et al.}, Phys. Rev. Lett. {\bf 81}, 3811 (1998).
\bibitem{dal}
For a review, see
F. Dalfovo, S. Giorgini, L.P. Pitaevskii, and S. Stringari, 
Rev. Mod. Phys. {\bf 71}, 463 (1999).
\bibitem{14} J.M. Kosterlitz and D.J. Thouless, J. Phys. C {\bf 6}, 1181 
(1973).
\bibitem{pop} V.N. Popov, \emph{Functional Integrals in Quantum Field Theory 
and Statistical Physics} (Riedel, Dordrecht, 1983); \emph{Functional Integrals 
and Collective Excitations} (Cambridge University Press, Cambridge, 1987).
\bibitem{fis} D.S. Fisher and P.C. Hohenberg, Phys. Rev. B {\bf 37}, 4936 
(1988).
\bibitem{tps2}
P. Pieri, G.C. Strinati, and I. Tifrea (in preparation). 
\bibitem{11} M. Schick, Phys. Rev. A {\bf 3}, 1067 (1971).
\bibitem{hin}
D.F. Hines, N.E. Frankel, and D.J. Mitchell, Phys. Lett. {\bf 68 A}, 12 
(1978).
\bibitem{kol} E.B. Kolomeisky and J.P. Straley, Phys. Rev. B {\bf 46}, 11749
(1992).
\bibitem{lie}
E.H. Lieb and J. Yngvason, math-ph/0002014, J. Stat. Phys. (in press).
\bibitem{cha} C. Chang and R. Friedberg, Phys. Rev. B {\bf 51}, 1117 (1995).
\bibitem{gru} P. Gr\"uter, D. Ceperley, and F. Lalo\"e, Phys. Rev. Lett. 
{\bf 79}, 3549 (1997).
\bibitem{hol} M. Holzmann and W. Krauth, Phys. Rev. Lett. {\bf 83}, 2687 
(1999).
\bibitem{may}
R.M. May, Phys. Rev. {\bf 115}, 254 (1959).
\bibitem{bel} S.T. Beliaev, JETP {\bf 34}, 417 (1958) [Sov. Phys. JETP 
{\bf 7}, 289 (1958)].
\bibitem{sym} The interaction vertices represented by points in the  
diagrams of Fig. 1 are meant to be symmetrized:  $v(\bq_1,\bq_2,\bq_3,
\bq_4)=
v(\bq_1-\bq_3)+v(\bq_1-\bq_4)$. By expressing the self-energy diagrams of 
Fig.~1(a) in terms of unsymmetrized interaction vertices and taking 
into account the symmetry factor of these diagrams, Eqs.\rf{sel} 
and\rf{tma} follow.   
\bibitem{asy}
Throughout this paper, we adopted the same notation of Ref.~\onlinecite{fis},
with the symbol $\approx$ meaning ``asymptotically equal''.
\bibitem{mub}
In two dimensions, the chemical potential of the ideal Bose gas can be
evaluated analytically for {\em all}\/ temperatures and densities, yielding 
$
\mu_0(n,T)=T \ln(1-e^{-2 \pi n/m T}) 
$
[cf.~Ref.~\onlinecite{may}].
\bibitem{bay2} 
G. Baym, J.-P. Blaizot, and J. Zinn-Justin, 
Europhys. Lett. {\bf 49}, 150 (2000).
\bibitem{hua}
K. Huang, Phys. Rev. Lett. {\bf 83}, 3770 (1999).
\bibitem{sto1}
H.T.C. Stoof, Phys. Rev. Lett. {\bf 66}, 3148 (1991).
\bibitem{lee}
T.D. Lee and C.N. Yang, Phys. Rev. {\bf 112}, 1419 (1958).
\bibitem{sto}
H.T.C. Stoof, Phys. Rev. A {\bf 45}, 8398 (1992).
\bibitem{bay} 
G. Baym, J.-P. Blaizot, M. Holzmann, F. Lal\"oe, and D. Vautherin, 
Phys. Rev. Lett. {\bf 83}, 3770 (1999).
\bibitem{2} H. Ding \emph{et al.\/}, Nature {\bf 382}, 51 (1996); 
Phys. Rev. Lett. {\bf 78}, 2628 (1997).
\bibitem{3} A. G. Loeser \emph{et al. \/}, Science {\bf 273}, 325 (1996).
\bibitem{1} Y.J. Uemura \emph{et al.\/}, Phys. Rev. Lett. {\bf 62}, 
2317 (1989).
\bibitem{7} R. Haussmann, Z. Phys. B {\bf 91}, 291 (1993).
\bibitem{8} P. Pieri and G.C. Strinati, Phys. Rev. B. {\bf 61}, 15370 (2000).
\bibitem{5} M. Randeria, Ji-Min Duan, and Lih-Yir Shieh, Phys. Rev. Lett. 
{\bf 62}, 981 (1989); Phys. Rev. B {\bf 41}, 327 (1990).
\bibitem{6} F. Pistolesi and G. C. Strinati, Phys. Rev. B 
{\bf 53}, 15168 (1996).
\bibitem{neg} J.W. Negele and H. Orland, {\em Quantum Many-Particle Systems}
(Addison-Wesley, New York, 1988), section 5.3. 

\end{references}
\end{document}